%% file: LowMass.tex
\documentclass[aps,prl,groupedaddress,superscriptaddress,letterpaper,twocolumn]{revtex4-2}

\pdfoutput=1

\usepackage[utf8]{inputenc}
\usepackage[T1]{fontenc}
\usepackage{ae}
\usepackage{aecompl}
\usepackage{hyperref}
\usepackage{xcolor}
\usepackage{graphicx}
\usepackage{float}
\usepackage{amsmath,amssymb}
\usepackage{multirow} 
\usepackage{tabularx}
\usepackage{xspace}

\newcommand{\DM}{\mbox{DM}\xspace}
\newcommand{\DSfs}{\mbox{DS-50}\xspace}
\newcommand{\DStw}{\mbox{DS-20k}\xspace}
\newcommand{\NR}{\mbox{NR}}
\newcommand{\ER}{\mbox{ER}}
\newcommand{\Ne}{\mbox{$N_{e^-}$}}
\newcommand{\DmFormFactor}{\ensuremath{F_{\mathrm{DM}}}\xspace}

\begin{document}
\title{DarkSide-20k sensitivity to light dark matter particles}
\input{authors}
\collaboration{The DarkSide-20k Collaboration}\thanks{ds-ed@lists.infn.it}


\begin{abstract}
\begin{center}
\textbf{Abstract}
\end{center}
The dual-phase liquid argon time projection chamber is presently one of the leading technologies to search for dark matter particles with masses below 10~GeV/c$^2$. This was demonstrated by the DarkSide-50 experiment with approximately 50~kg of low-radioactivity liquid argon as target material. The next generation experiment DarkSide-20k, currently under construction, will use 1,000~times more argon and is expected to start operation in 2027. Based on the DarkSide-50 experience, here we assess the DarkSide-20k sensitivity to models predicting light dark matter particles, including Weakly Interacting Massive Particles (WIMPs) and sub-GeV/c$^2$ particles interacting with electrons in argon atoms. With one year of data, a sensitivity improvement to dark matter interaction cross-sections by at least one order of magnitude with respect to DarkSide-50 is expected for all these models. A sensitivity to WIMP--nucleon interaction cross-sections below $1\times10^{-42}$~cm$^2$ is achievable for WIMP masses above 800~MeV/c$^2$.
With 10 years exposure, the neutrino fog can be reached for WIMP masses around 5~GeV/c$^2$.
\end{abstract}
\maketitle

\section{Introduction}
The presence of dark matter (\DM) in the universe is supported by many observations based on gravitational effects~\cite{Rubin:1980zd,Clowe:2006eq,Planck:2018vyg} but its real nature remains unknown. Dark matter may consist of an undiscovered elementary particle~\cite{Bertone:2004pz}. A leading
candidate is a Weakly Interacting Massive
Particle (WIMP), with a mass ranging from 10~GeV/c$^2$ to few TeV/c$^2$. This range is extensively searched for via elastic scattering off atomic nuclei -- later called nuclear recoils (\NR). These searches often use underground Time Projection chambers (TPC)~\cite{PandaX-4T:2021bab, LZ:2022lsv, XENON:2023cxc, DarkSide:2018kuk, DEAP:2019yzn}. The lack of concrete evidence of direct \DM detection so far motivates the search for lighter WIMPs, below 10~GeV/c$^2$, and for light DM candidates interacting with shell electrons -- later called electron recoils (\ER) -- which may subsequently produce sufficiently large ionization signals~\cite{Essig:2011nj}.

DarkSide-20k (\DStw) is the next generation of liquid argon (LAr) dual-phase TPCs, presently in construction at INFN Laboratori Nazionali del Gran Sasso in Italy. It is expected to start taking data in 2027. It is primarily designed to perform a nearly instrumental background-free search for high mass ($>$10~GeV/c$^2$) WIMPs. \DStw aims for $<0.1$ background events with an exposure of 200~ton$\cdot$year. The detection mechanism relies on the combined observation of the  scintillation light (S1) and ionization (S2) signals. DarkSide-50 (\DSfs), a first generation LAr dual-phase TPC, already demonstrated the capability of the technology and obtained world best sensitivities to light DM particles using only the S2 signal, when the S1 signal is no longer observable~\cite{DarkSide:2018bpj,DarkSide:2018ppu}. The analysis was based solely on the number of electrons in S2 (\Ne) as discriminating variable. An update based on the capability to measure the liquid argon ionization yield for low energy electron and nuclear recoils down to $\sim$~180 eVer and $\sim$~500 eVnr~\cite{DarkSide:2021bnz}, respectively, was one of the key ingredients to obtain world best limits for WIMPs in the mass range [1.2, 3.6]~GeV/c$^2$~\cite{DarkSide-50:2022qzh}. A factor of 10 improvement with respect to the previously published limit of Ref.~\cite{DarkSide:2018bpj} was achieved. The analysis was also used to place limits on DM--nucleon interaction via the Migdal effect~\cite{DarkSide:2022dhx} and on sub-GeV/c$^2$ DM--electron scattering~\cite{DarkSide:2022knj}. 
The stability of the electroluminescence yield has been measured to be better than 0.5\% over almost three years~\cite{DarkSide-50:2023nes}.
Based on these successes of \DSfs, this article presents the low-mass DM sensitivity prospects for the \DStw detector. \DStw will increase the low-radioactivity LAr volume by about a factor of~1,000 with respect to \DSfs. It will also significantly improve the radio-purity of the components surrounding the active volume. A detector specifically designed for the investigation of light dark matter using LAr and assuming further isotopic depletion via cryogenic distillation has also been recently studied~\cite{GlobalArgonDarkMatter:2022ppc}.

\section{The \DStw detector}

The \DStw TPC is filled with a 49.7~ton active mass of argon extracted from underground CO$_2$ wells, hence called Underground Argon (UAr). The TPC is shaped as a prism with an octagonal base, with a vertical drift length of 348~cm and an octagonal inscribed circle diameter of 350~cm. 
The active volume is immersed in a uniform electric field generated by applying a voltage potential of 73.4~kV between the anode and the cathode made of transparent acrylic (PMMA) coated with a conductive material (Clevios\texttrademark). The corresponding maximum drift time for the ionization electrons is 3.7~ms. A set of 200~$\mu$m wide stainless steel wires spaced by 3~mm located 3~mm below the liquid level is used to define a high field extraction region for drifting electrons. A multiplication region filled with gaseous argon, 7~mm thick between the liquid level and the anode, allows the S2 electroluminescence signal to develop. 

\begin{figure}[th]
\centering
\includegraphics[width=\columnwidth]{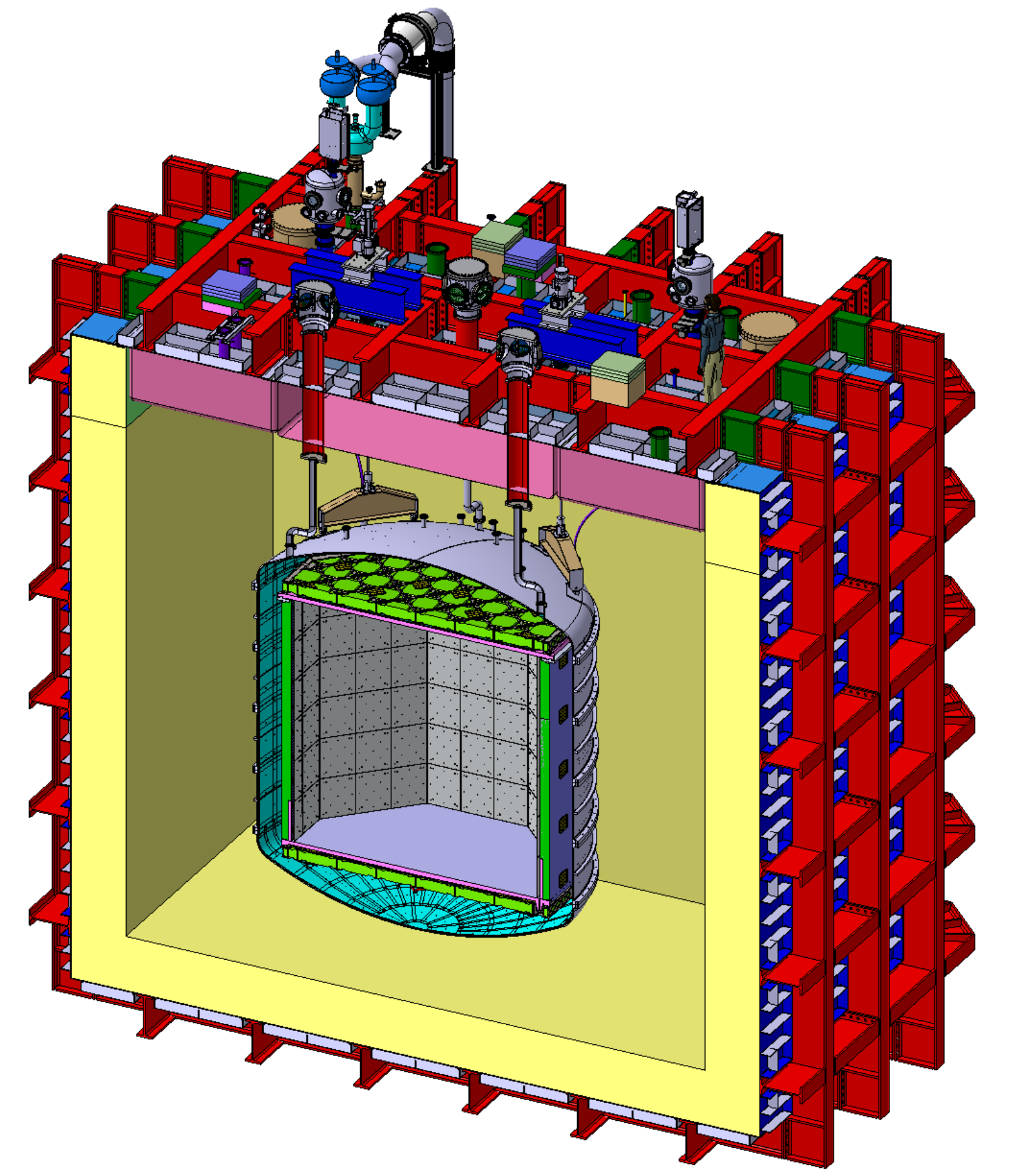}
\caption{Cross-section of the \DStw detector showing 
at its center the TPC, with its acrylic walls in green and electrodes in pink. The stainless steel vessel surrounding it is shown in gray, immersed in the  ProtoDUNE-like cryostat.
}
\label{fig:Detector}
\end{figure}

To ensure the best possible collection of both scintillation and ionization-induced photons, the inner TPC walls are covered with reflectors. All the inner surfaces are coated with tetraphenyl butadiene to shift the native argon UV light to visible wavelength for which photo-detection efficiency is maximal. Two planes of cryogenic Silicon Photo Multipliers (SiPMs) covering the top and the bottom faces of the TPC detect the light signals. 
The $\sim$200k SiPMs are gathered in  5$\times$5~cm$^2$ arrays, called Photo Detector Modules (PDMs)~\cite{DIncecco:2017bau}. PDMs include the front-end electronics~\cite{DIncecco:2017qta} and are arranged in~2112 readout channels. 
The S2 yield is about 25 photo-electrons per ionization electron  extracted in the gas pocket. If neutrons scatter in the TPC and produce a WIMP-like signal, they are likely to be captured in the 15~cm thick acrylic TPC walls or endcaps that enclose the SiPM planes, with the latter possibly further doped with Gadolinium. Neutrons captured in this way release $\gamma$-rays which are detected in the TPC and/or the 32~ton UAr veto surrounding it, which is read by 480 photo detector channels. The TPC and this UAr veto are housed in a stainless steel (SS) vessel. This SS vessel is immersed in a bath of 650~tons of atmospheric argon (AAr), acting as a shield and an outer veto detector for muons and associated products. The AAr is contained in a ProtoDUNE-like membrane cryostat~\cite{DUNE:2020lwj}. 
\DStw\ adopts a triggerless data acquisition system ensuring to save all SiPM signals with close to 100\% efficiency for single photo-electrons.
Figure~\ref{fig:Detector} shows a cross-section of \DStw. 

\section{Selection}

The analysis relies on the strategy successfully put in place for \DSfs~\cite{DarkSide-50:2022qzh} and adapted to \DStw. WIMPs are expected to scatter 
only once in the LAr volume of the TPC. As each particle interaction is associated with a S2 pulse, events with a single S2 pulse are selected. Pulses are required to be isolated from any other S2 pulse preceding or following by more than 3.7 ms -- which corresponds to the maximum electron drift time. 
It is assumed that two close S2 pulses can be distinguished if they are separated in time by more than 2~$\mu$s (2~mm along the drift direction), as achieved in \DSfs.
Relaxing this assumption produces a negligible impact on the expected sensitivity.
Anomalously low S2 could come from $\alpha$-background. It is assumed that S2 signals can be identified without introducing significant inefficiencies, and that surface $\alpha$-background that may produce low energy S2 can be efficiently suppressed, as in \DSfs.
Before any selection procedure, the total rate of ER-events from electron and $\gamma$-ray backgrounds from radioactive decays is estimated to be 80~Hz (0.0016~Hz/kg of UAr) in the TPC, to be compared with 1.5~Hz in \DSfs (0.03~Hz/kg of UAr). The  accidental coincidences within the 3.7~ms isolation window between the S2 pulses induced by this background and by the signal result in an effective livetime of 51\%.
Unlike the high mass WIMP search analysis, it is not possible to use the pulse shape discrimination of the S1 signal~\cite{DEAP-3600:2017uua,DEAP:2019yzn} to remove the ER background. Instead, the selection aims at mitigating $\gamma$-rays and X-rays from radioactive isotopes of the detector components surrounding the UAr active volume (including PDMs). They are efficiently suppressed by a fiducialization based on the S2 pattern in the transverse direction of the TPC, removing an outer 30~cm thick octagonal shell. The position reconstruction resolution along this direction is estimated to be better than 3 cm. As the present analysis is based solely on S2, the drift time, computed as the time difference between S1 and S2, is not available and no fiducialization is performed along the electron drift direction. This procedure leads to a UAr fiducial mass of 34.2~tons, hence an exposure of 17.4~ton$\cdot$year for one year of data, taking into the account the effective livetime. 

\section{Detector response model}

The number of electrons in S2 is derived from the energy deposited by a single scatter event in the UAr using the ER and NR ionization yields measured by \DSfs~\cite{DarkSide:2021bnz}. The intrinsic fluctuations to ER signals are modelled with an empirical fudge factor implemented as a Fano factor measured by \DSfs~\cite{DarkSide:2021bnz}. Fluctuations from the ionization quenching effect in NR are not known and two models, assuming no fluctuation (NQ) or binomial fluctuations between detectable and undetectable quanta (QF), are considered~\cite{DarkSide-50:2022qzh}. Unless explicitly stated, QF is assumed in this article. 
A Geant4-based Monte Carlo package~\cite{DarkSide:2017wdu} is used for an accurate simulation of light production, propagation and detection for background and signal events. Additionally, effects due to SiPM and electronics noise are simulated on the basis of pre-production SiPM module performance, resulting in a 23\% single electron response resolution. Electron losses during the drift due to attachment to impurities are taken into account assuming a 16~ms lifetime, as measured in \DSfs~\cite{DarkSide:2017wdu}.

\section{Background model}

The $\beta$-decay background intrinsic to LAr is fully dominated by the radioactive isotopes $^{39}$Ar and $^{85}$Kr present in the active volume of UAr. Extracting the argon from underground (same source as \DSfs) significantly reduces their contamination with respect to AAr~\cite{DarkSide:2018kuk}. Their activities are assumed to be 0.73~mBq$/$kg and 1.9$\times10^{-2}$~mBq$/$kg, respectively. The former has the same level as in \DSfs while the latter is reduced by a factor 100 compared to the one of \DSfs, thanks to a new multiple distillation column system that has been added at the UAr extraction plant. The argon and krypton spectral shapes
are based on calculations of atomic exchange and screening effects, validated on measured $^{63}$Ni and $^{241}$Pu spectra with a 200~eV threshold~\cite{Haselschwardt:2020iey,PhysRevA.90.012501}.
Below this value, a linearly increasing uncertainty on the corresponding corrections is assumed, reaching 25\% at 0~eV~\cite{DarkSide-50:2022qzh}.
Further systematics on the spectral shape originate from the uncertainty on the $Q$-value (1\% for $^{39}$Ar and 0.4\% for $^{85}$Kr). 

External $\gamma$-ray and X-ray backgrounds come from materials used to build the whole inner detector system. Materials are carefully selected for low levels of radioactivity and their activities are measured in an extensive material screening campaign based on the combination of different radio assay techniques. As a summary, Table~\ref{tab:bg} lists the expected activity of each considered radio-contaminant of the inner detector materials that may lead to $\gamma$-ray and X-ray background. Each isotope is simulated uniformly in the component material and decaying particles are tracked over the \DStw geometry. Thanks to the use of SiPMs instead of classical photomultipliers, the use of PMMA walls for the TPC and the minimization of the amount of passive material, the $\gamma$-ray background level is expected to be reduced with respect to \DSfs by a factor 2.5 per surface area orthogonal to the electron drift direction.
This validates the extreme care taken to consider radio-pure materials in the design of the \DStw inner detector.

\begin{table}[h]{
\begin{tabular}{| l | c | c | c | c |}
\hline
 Radio-       & \multicolumn{3}{c|}{Activity (Bq)}  \\
 \cline{2-4}
 contaminant      & TPC  & PDMs  &  SS vessel   \\
 \hline
 $^{238}$U up   & 16.1 & 38.8   & 21  \\
 $^{238}$U mid  & 11.5 & 18.4   & 8.8 \\
 $^{238}$U low  & 16.4 & 449   & 62  \\
 $^{232}$Th     & 4.2  & 17.8   & 33  \\
 $^{235}$U      & 0.7 & 1.8   & 1.0 \\
 $^{137}$Cs     & 2.5 & 2.9  & 5.0 \\
 $^{60}$Co      & 2.0 & 5.1   & 13\\
 $^{40}$K       & 102 & 269   & 49 \\
\hline
\end{tabular}
\caption{Estimation of \DStw material activities for each radio-contaminant, in Bq, for external $\gamma$-ray and X-ray background sources. The activity measurements are reported for chain progenitors only. In the $^{238}$U decay chain, up covers from $^{238}$U to $^{222}$Rn, mid from $^{222}$Rn to $^{210}$Pb and low from $^{210}$Pb to $^{206}$Pb.}
\label{tab:bg}
}
\end{table}

``Spurious'' electrons (SE), whose origin might be ionization electrons trapped by impurities and released later, are a major component of the background at low number of electrons (\Ne$<4$). 
The ionization spectrum model is built by fitting \DSfs data in this \Ne\ range, assuming the same spectrum in \DStw, with a Poisson distribution convolved with a Gaussian accounting for the single electron response. 
The origin of the spurious electron signals is assumed to be completely explained by the impurity mechanism in UAr, with the same level of impurities as in \DSfs. The expected rate in \DStw is extrapolated from the \DSfs rate as in Ref.~\cite{GlobalArgonDarkMatter:2022ppc}.
This background will need to be thoroughly characterized once real \DStw\ data is available.
The systematic uncertainty on the spectral shape derived from the fit to \DSfs data is assigned to the \DStw SE modelling.

Other backgrounds to the DM search coming from the interaction of neutrinos via coherent elastic scattering off nucleus (CE$\nu$NS)~\cite{Cadeddu:2023tkp}, recently observed with argon~\cite{COHERENT:2020iec} and xenon~\cite{XENON:2024ijk,PandaX:2024muv} nuclei, have been considered. The study includes radiative corrections~\cite{AtzoriCorona:2024rtv} and an accurate parametrization of the nuclear structure~\cite{Cadeddu:2020lky}. The main contribution impacting this analysis comes from neutrinos from solar $^{8}$B which deposit less than 10~keV in UAr. Elastic scattering ($\nu$-ES) off argon electrons~\cite{AtzoriCorona:2022qrf} have also been considered, surpassing CE$\nu$NS for \Ne$>$30. In both cases, neutrino fluxes are normalized according to Ref.~\cite{Baxter:2021pqo}. 
Finally, the rate of NRs from radiogenic and cosmogenic neutrons is expected to be over four orders of magnitude lower than the ER one, and therefore not considered in this analysis. 

\section{Signal models}

The signal models are derived assuming the standard isothermal WIMP halo model with an escape velocity of 544~km/s, a local standard of rest velocity of 238~km/s, and a local dark matter density of 0.3~GeV/c$^2$/cm$^3$~\cite{Baxter:2021pqo}.
WIMPs are assumed to elastically scatter off nucleons. 
Atomic effects predicted by Migdal~\cite{Migdal:1941} could add
an extra emission of electrons to a fraction of nuclear recoils, increasing the sensitivity to low mass WIMPs. 
As experimental efforts to confirm the existence of such effects in nuclear scattering are still underway~\cite{Araujo:2022wjh,Xu:2023wev,Bell:2021ihi},
WIMP signals with and without them, modelled as in Ref.~\cite{DarkSide:2022dhx}, are considered in this article.

Other dark matter candidates interacting with electrons and producing a S2 signal are possible. First, fermion or scalar boson light dark matter particles with a mass below 1~GeV/c$^2$ could interact with bound electrons via a vector mediator. The interaction and cross-section depend on the momentum-transfer $q$ via a form factor \DmFormFactor and two benchmark models are considered: a heavy mediator with \DmFormFactor$=1$ and a light mediator with \DmFormFactor$=(\alpha m_e/q)^2$ where $\alpha$ is the fine structure constant and $m_e$ the mass of the electron. Other possibilities are the absorption of axion-like particles -- coupled to electron via $g_{Ae}$, and vector-boson like dark photon -- mixing with photon via the $\kappa$ parameter -- by argon shell electrons. Finally sterile neutrinos, mixing with an active neutrino state by an angle $|U_{e4}|^2$, could inelastically scatter off a bound electron. All these models are described in details in Ref.~\cite{DarkSide:2022knj}.

\begin{table}[hbtp]
\begin{tabular}{|l|l|l|}
\hline
   &   Source uncertainty & Affected      \\
   &   & components      \\
\hline
\hline
\multirow{5}{*}{\rotatebox[origin=c]{90}{Amplitude}}    
&  5\% on the exposure & All \\ 
\cline{2-3} 
& 15\% on $^{39}$Ar activity  & $^{39}$Ar \\
\cline{2-3} 
& 15\% on $^{85}$Kr activity  & $^{85}$Kr \\
\cline{2-3} 
& 20\% on SE normalization & SE \\
\cline{2-3} 
& 10\% on activity from PDMs  & PDMs \\
\cline{2-3} 
& 10\%  on activity from the vessel    & Vessel \\
\cline{2-3} 
& 10\%  on activity from the TPC    & TPC \\ \cline{2-3} 
& 10\%  on neutrinos normalization  & Neutrinos \\ 
\hline
\hline
\multirow{6}{*}{\rotatebox[origin=c]{90}{Shape}}    
& atomic exchange and screening & $^{39}$Ar \\
\cline{2-3} 
& atomic exchange and screening & $^{85}$Kr \\
\cline{2-3} 
& 1\% on the $^{39}$Ar-decay $Q$-value &  $^{39}$Ar \\
\cline{2-3} 
& 0.4\% on the $^{85}$Kr-decay $Q$-value &  $^{85}$Kr \\
\cline{2-3} 
& SE modelling  &  SE \\
\cline{2-3} 
& ER ionization response &  All backgrounds\\
&  &  but CE$\nu$NS, SE\\
\cline{2-3} 
& NR ionization response & WIMP, CE$\nu$NS\\
\hline 
\end{tabular}
\caption{List of systematic uncertainties, their sources, and impacted signal and background components included in the binned profile likelihood. Any considered spectrum is equally affected by the uncertainty on the dataset exposure, but differs on the ionization response on the basis of the recoil type. The pre-fit uncertainty values are adapted from \DSfs\ analysis~\cite{DarkSide-50:2022qzh}.}
\label{tab:DS20ksys}
\end{table}

\section{Results and Discussion}

Using these signal and background models, \DStw prospects for 90\% C.L. exclusion limits are derived from a binned profile-likelihood fit implemented in the RooFit package~\cite{Cranmer:1456844}. The likelihood is built as the product of Poissonian terms, one for each of the considered
bins. Systematic uncertainties that are believed to be reachable at the time of data taking
are accounted for by introducing the nuisance parameters shown in Table~\ref{tab:DS20ksys}. They are classified as amplitude or shape systematics, the latter accounting for uncertainties on $^{39}$Ar and $^{85}$Kr $\beta$-decays~\cite{DarkSide-50:2022qzh} and on spurious electrons modelling, as well as for spectral distortions from the ionization response. 

\begin{figure}[htbp]
\centering
\includegraphics[width=\columnwidth]{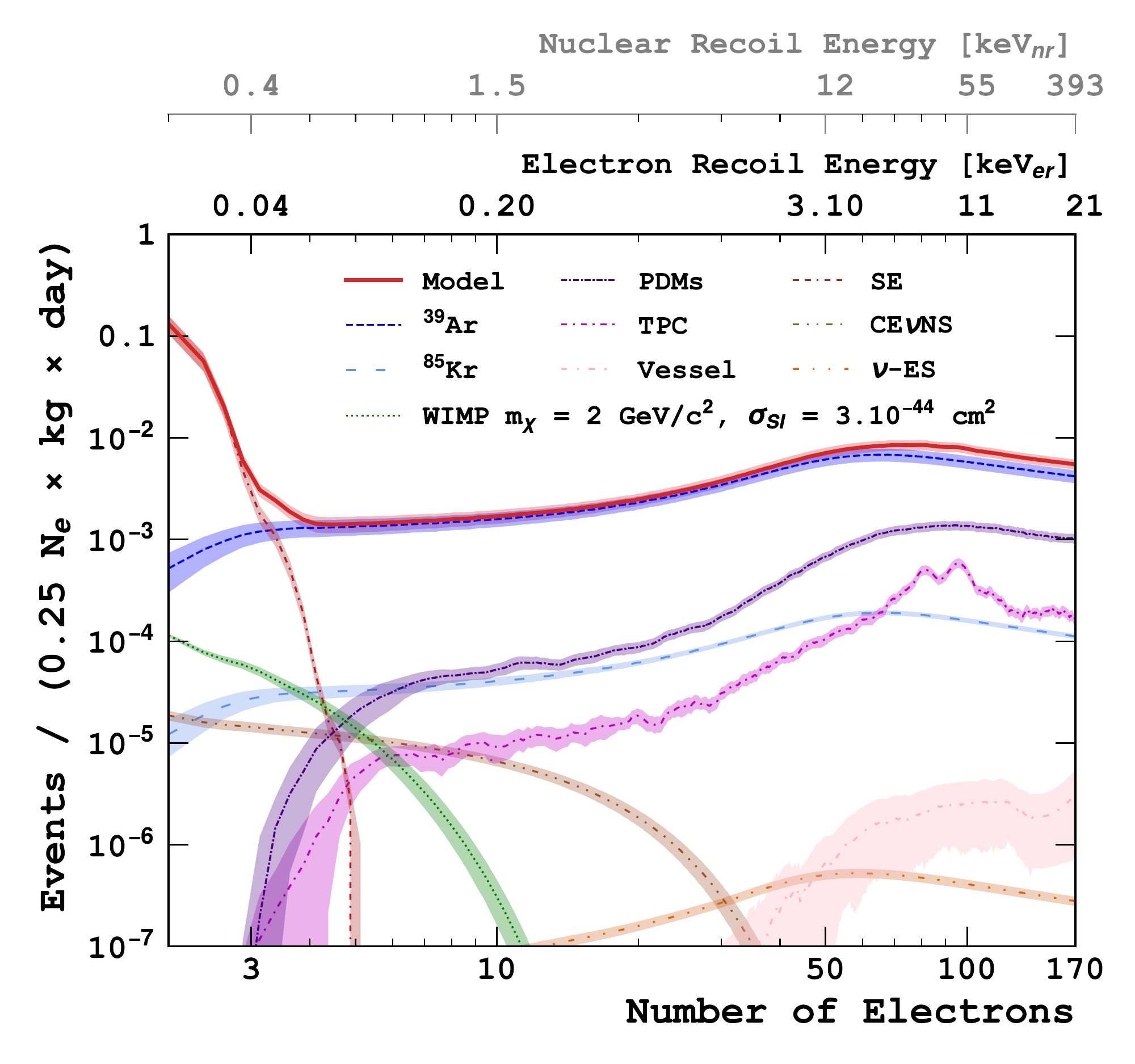}
\caption{Pre-fit \DStw \Ne\ spectra distribution. The corresponding ER and NR energy scales are indicated at the top. The pre-fit background model (red line) is shown with its uncertainties (shaded area). Contributions from all the components of the background are also shown. A typical signal model is superimposed for illustration (green dotted line), assuming a 2~GeV/c$^2$ WIMP mass with a cross-section of $3\times10^{-44}$ cm$^2$ and the QF scenario for the signal fluctuations. }
\label{fig:Sensitivity}
\end{figure}

Figure~\ref{fig:Sensitivity} shows the pre-fit \Ne\ distribution for the background model. The $^{39}$Ar component is dominating for \Ne$\geq4$. The SE contribution dominates in the $2\leq\Ne\leq3$ range and is a factor $\sim$18 below $^{39}$Ar at \Ne$=4$. The contribution from the PDMs is dominating the external $\gamma$-ray background.
The neutrino background from CE$\nu$NS and $\nu$-ES is two to four orders of magnitude below the $^{39}$Ar background. 
A typical signal, corresponding to a 2~GeV/c$^2$ WIMP mass with a cross-section of $3\times10^{-44}$~cm$^2$, is superimposed for illustration. 

In the following, given the unknowns on the SE background before data taking, two different fit strategies are envisaged: the first one (conservative) is reproducing the \DSfs approach and uses the \Ne\ range from 4 to 170, while the second one (ultimate) assumes a good understanding of the spurious electrons in \DStw and uses the total fit range from 2 to 170. Nuisance parameters affecting $^{39}$Ar, PDMs, TPC and spurious electrons (only in the ultimate fit case for the latter) are strongly constrained by the fit, since they are related to the dominant backgrounds. For the same reason, a strong (anti)-correlation exists between these amplitude nuisance parameters and the one associated to the exposure. 
As $^{39}$Ar is the dominant background, its spectral shape will need to be computed with higher precision in order to interpret any possible future excess in \DStw as a signal. 

\begin{figure*}[htbp]
\centering
\includegraphics[width=\textwidth]{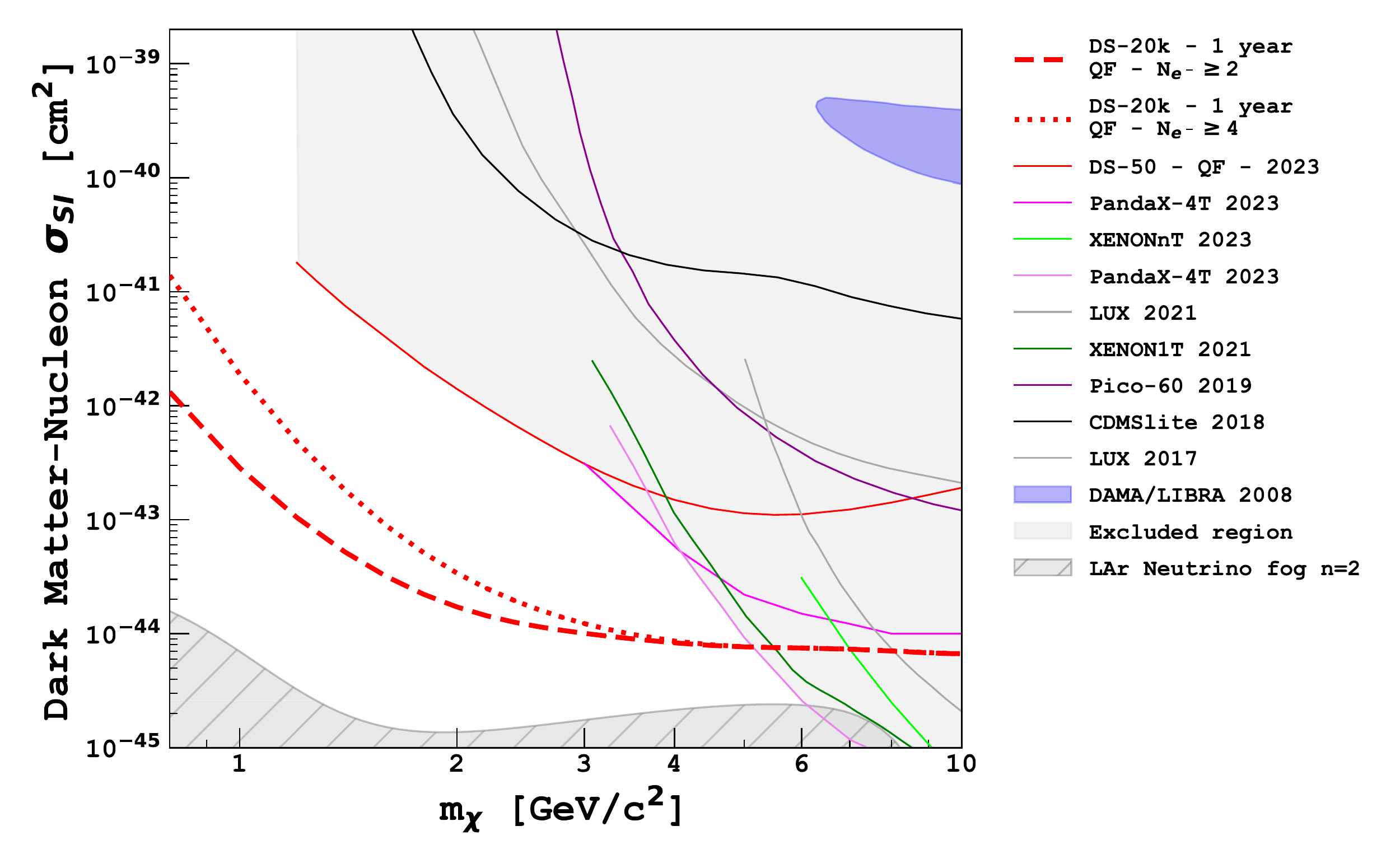}
\caption{Expected \DStw 90\% C.L. exclusion limits for spin-independent WIMP NR with quenching fluctuations (QF) are shown as bold red lines (dotted: fit from \Ne=4, dashed: fit from \Ne=2). One year of data is assumed. They are compared to the published 90\% C.L. limits from \DSfs~\cite{DarkSide-50:2022qzh} and from other experiments~\cite{XENON:2023cxc,PandaX:2022aac,PandaX:2022xqx,XENON:2020gfr,LUX:2020yym,LUX:2016ggv,SuperCDMS:2017nns,PICO:2019vsc}, with currently excluded parameter space shaded in light gray, as well as claimed discovery from Ref.~\cite{Bernabei:2013xsa}. The neutrino fog in LAr with index $n=2$~\cite{OHare:2021utq} is also shown. A local dark matter density of 0.3~GeV/c$^2$/cm$^3$ is assumed.} 
\label{fig:WIMP}
\end{figure*}

The \DStw expected median 90\% C.L. upper limits on spin-independent WIMP--nucleon cross-section ($\sigma_{SI}$), computed with the CLs prescription~\cite{Read:2002hq} and for the two fit strategies, are shown in Figure~\ref{fig:WIMP} for one year of data. They are compared to the published 90\% C.L. limits from \DSfs~\cite{DarkSide-50:2022qzh} and from other experiments~\cite{XENON:2023cxc,PandaX:2022aac,PandaX:2022xqx,XENON:2020gfr,LUX:2020yym,LUX:2016ggv,SuperCDMS:2017nns,PICO:2019vsc}. An improvement in sensitivity by up to a factor 40 over \DSfs is achieved using the conservative fit. This increases to a factor 170 at 1.2~GeV/c$^2$ WIMP mass with the ultimate fit, reaching a $\sigma_{SI}$ of $1\times10^{-43}$~cm$^2$. 
Overall, a sensitivity to $\sigma_{SI}$ below $1\times10^{-42}$~cm$^2$ is achieved for WIMP masses above 800~MeV/c$^2$, covering a large uncharted phase space with one year of data. The dominant systematic uncertainties come from the main background ($^{39}$Ar, SE) modelling and from the detector response (ER ionization yield). The sensitivity would be approximately 60\% better if systematics were neglected.
Results without quenching fluctuation on NR are shown in Supplementary Figures~1 and~2.

The sensitivity scales with the square root of the exposure. 
With 10 years exposure, the sensitivity will improve by a factor~3 compared to one year, whatever the WIMP mass: the neutrino fog in LAr with index $n=2$~\cite{OHare:2021utq} could be reached for WIMP masses around 5~GeV/c$^2$ (Supplementary Figure~3). 

To assess the robustness of these expectations, variations of detector response model and background activities with respect to the nominal assumptions have been considered. The electron lifetime, single electron response and $x-y$ resolutions and $^{39}$Ar activity have been varied by a factor 2 with respect to the nominal assumptions.
The main impact comes from the $^{39}$Ar activity, assumed to be the same as \DSfs, and from the single electron response resolution. If the latter is degraded by a factor~2 with respect to expectations, then the sensitivity would degrade at most by a factor~2 using the ultimate fit strategy, mainly affecting WIMP masses around 1~GeV/c$^2$. 
A significant improvement in sensitivity is expected if the UAr extraction plant would further reduce the contamination of $^{39}$Ar. For a factor 2 lower contamination, the sensitivity would improve up to a factor 1.8 
(Supplementary Figure~4).

\begin{figure*}[th]
\centering
\includegraphics[width=\columnwidth]{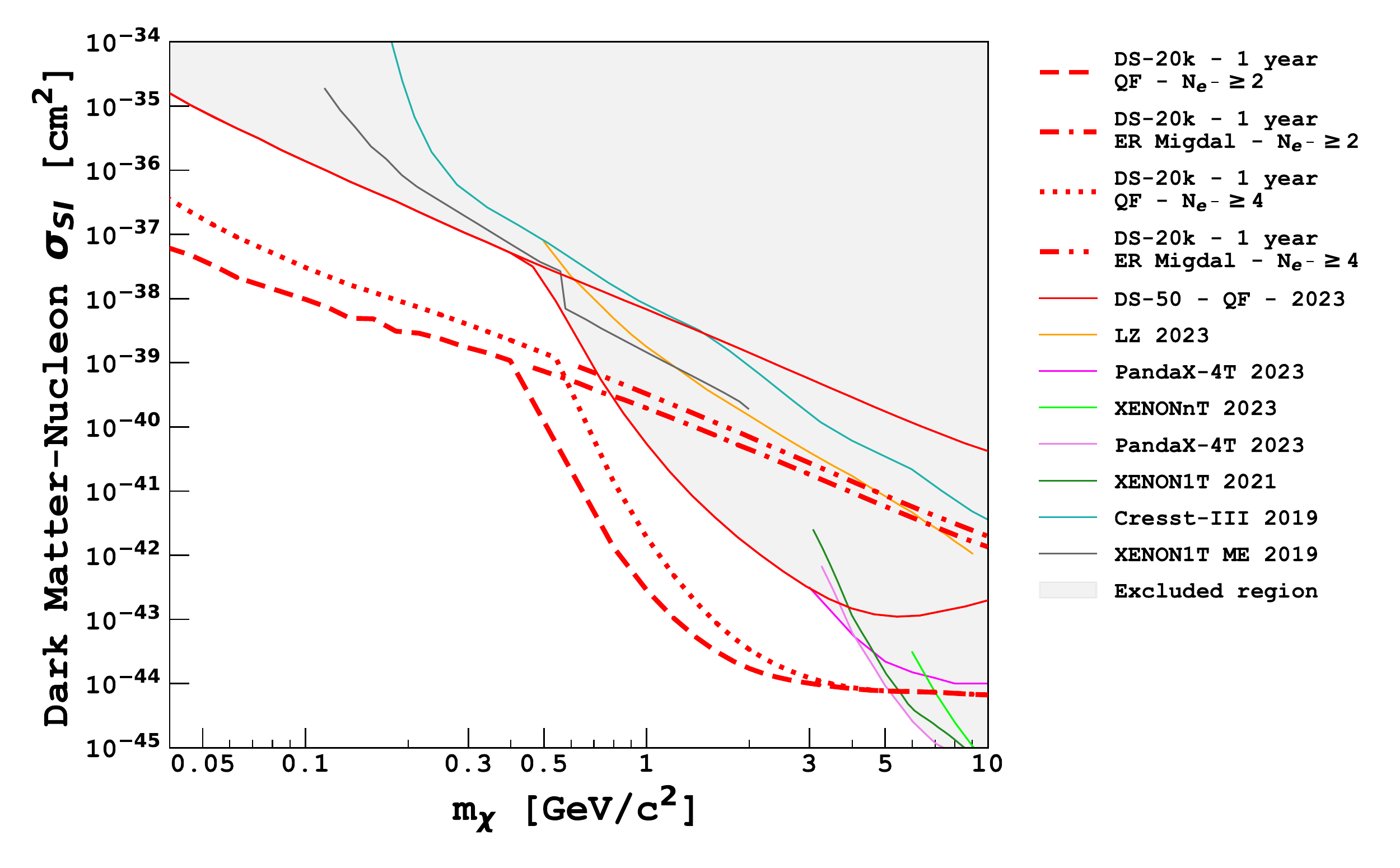}
\includegraphics[width=\columnwidth]{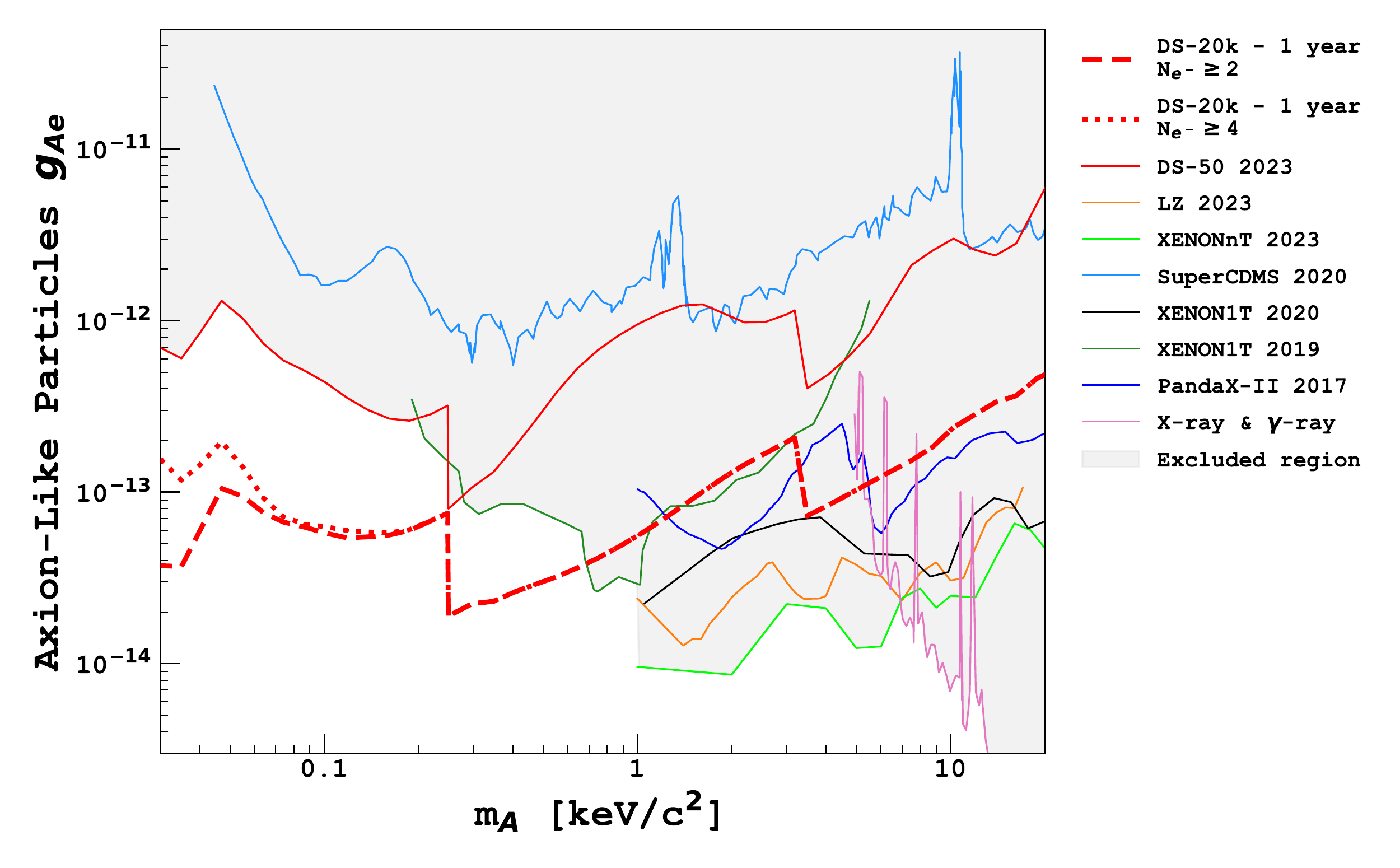}
\includegraphics[width=\columnwidth]{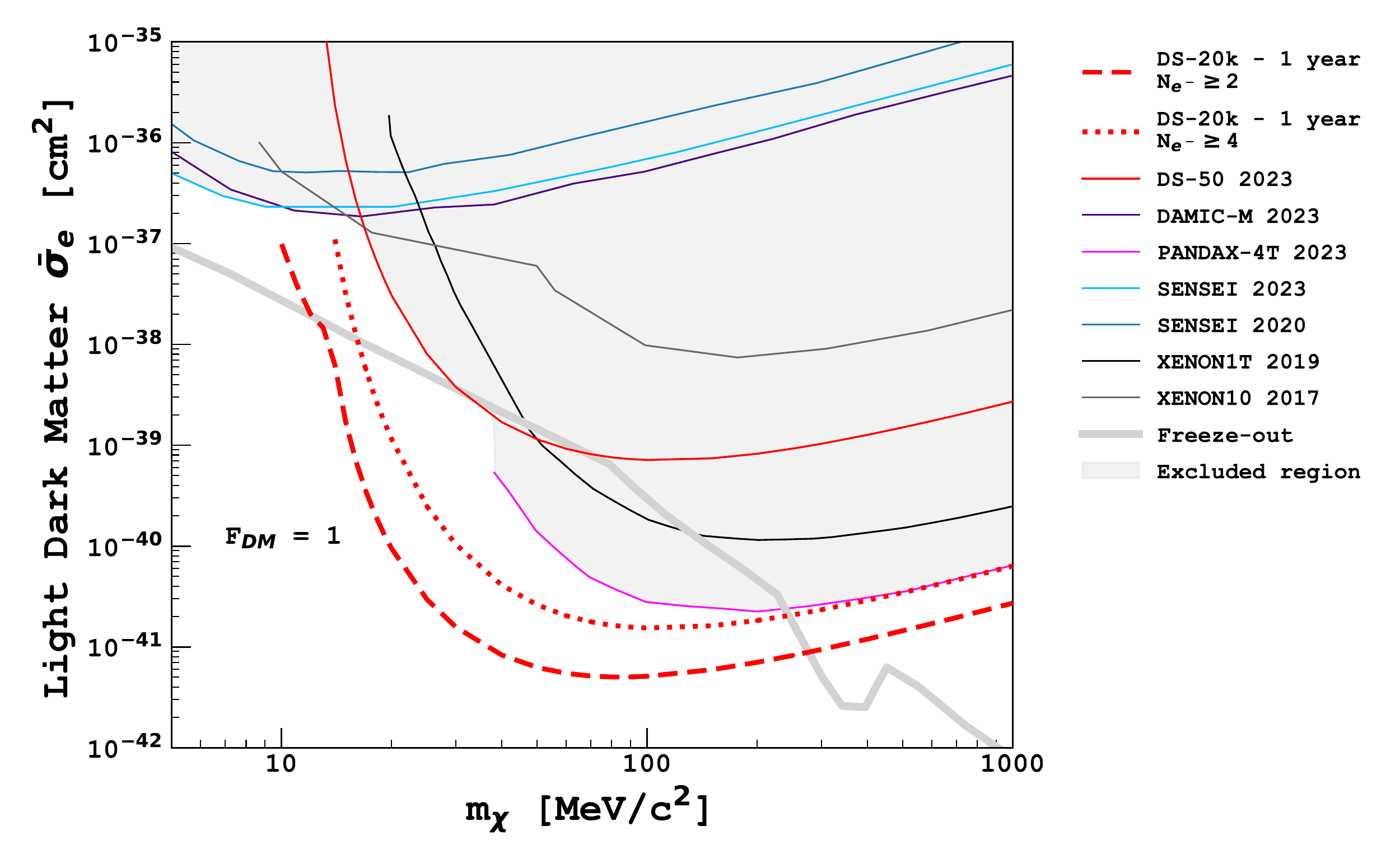}
\includegraphics[width=\columnwidth]{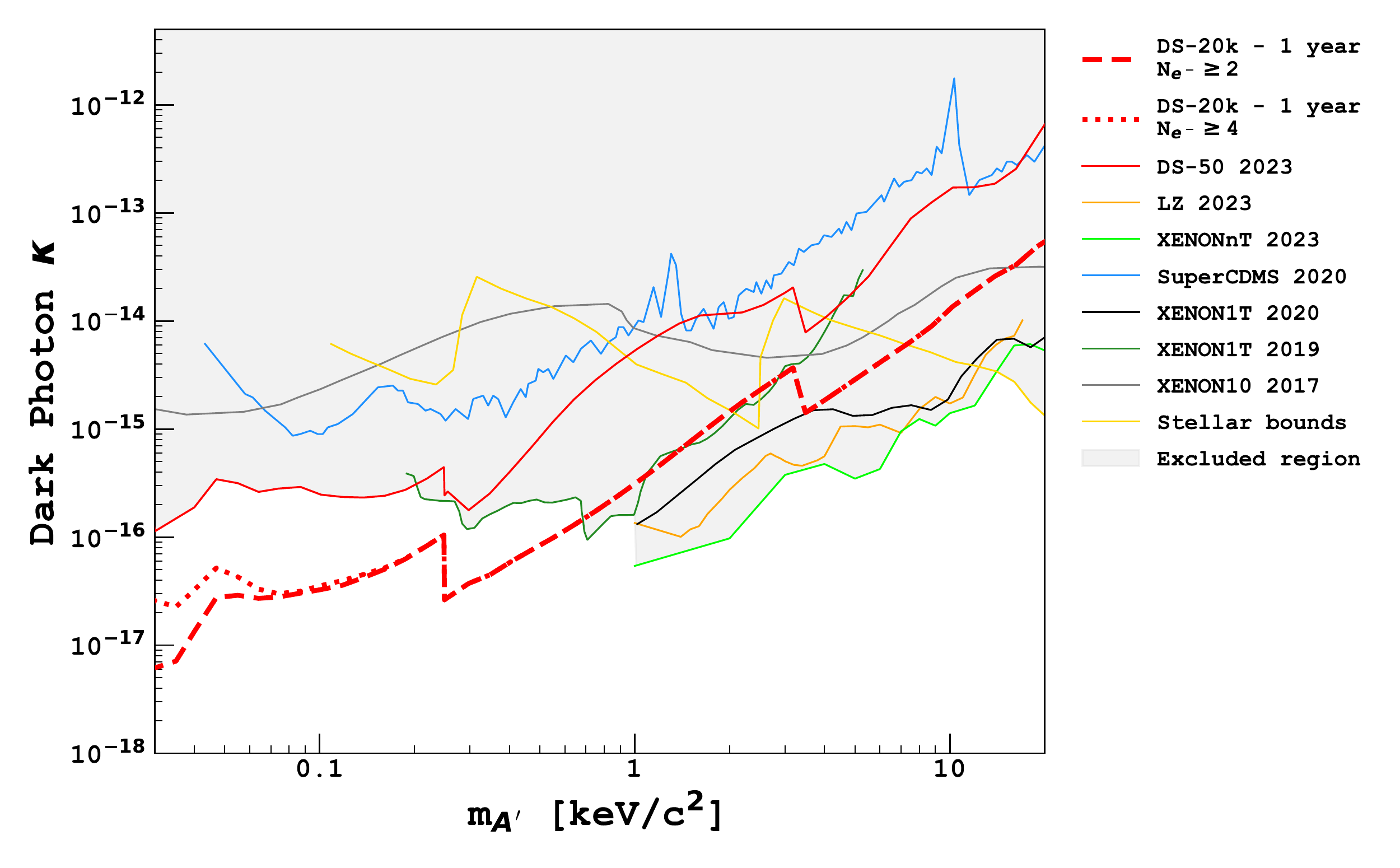}
\includegraphics[width=\columnwidth]{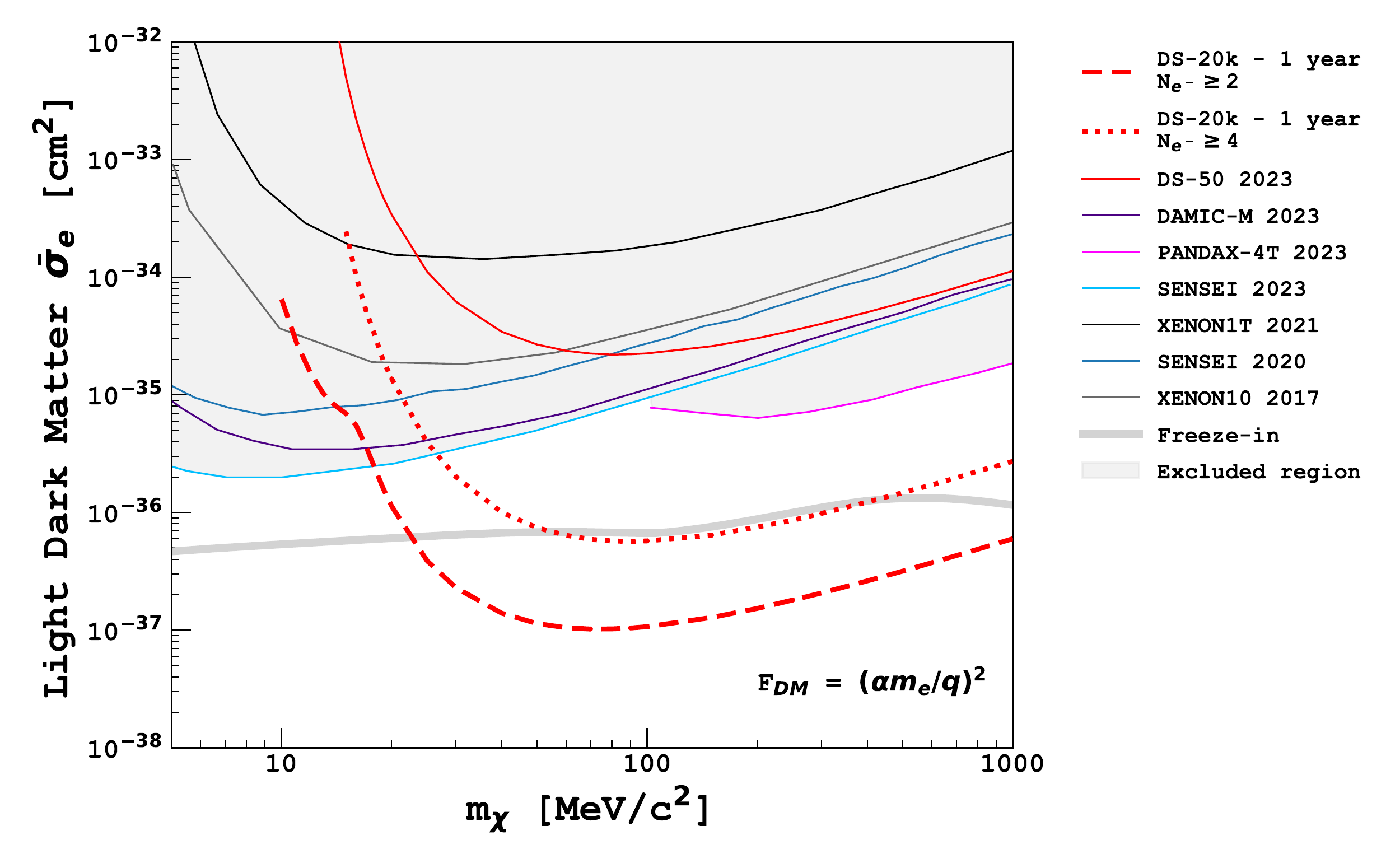}
\includegraphics[width=\columnwidth]{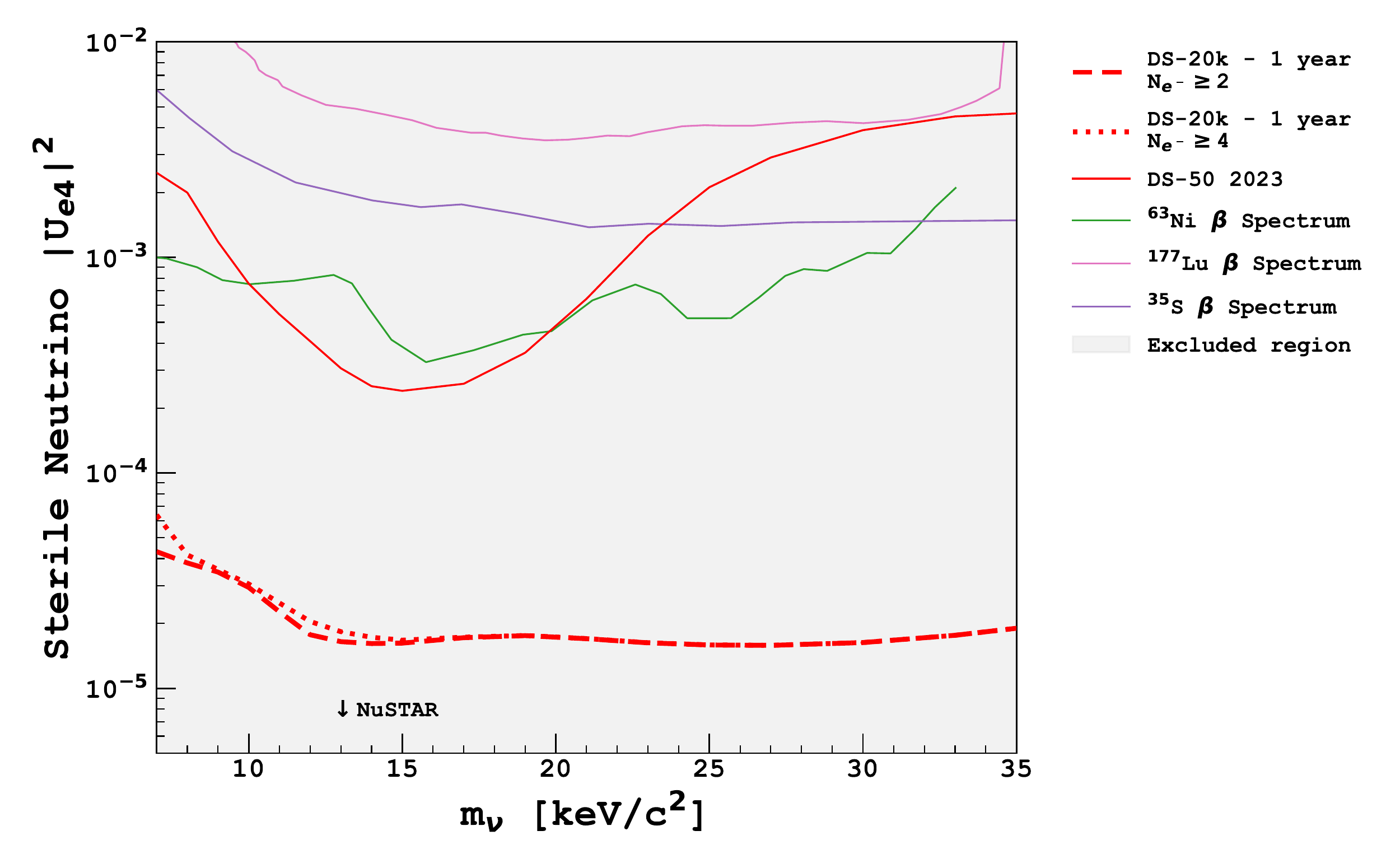}
\caption{Expected \DStw upper limits at 90\% C.L. for various signal models (bold red lines, dotted: fit from \Ne=4, dashed: fit from \Ne=2). The x-axis shows the mass of the
candidate while the y-axis shows the model parameter. One year of data is assumed. These results are compared
to the published 90\% C.L. limits from \DSfs~\cite{DarkSide:2022dhx,DarkSide:2022knj}, other experiments~\cite{XENON:2023cxc,PandaX:2022aac,PandaX:2022xqx,XENON:2019zpr,XENON:2020gfr,PhysRevD.100.102002,PhysRevLett.125.171802,PandaX-II:2021nsg,Essig:2017kqs,Xenon1tLightDM,XENON:2021qze,LZ:2023poo,DAMIC-M:2023gxo,SENSEI:2023zdf,HolzschuhSearchHeavyNeutrinos1999,MortaraEvidence17KeV1993,SchonertExperimentalLimitsHeavy1996,PhysRevLett.119.181806,SuperCDMS:2019jxx,AnDirectDetectionConstraints2015,AprileExcessElectronicRecoil2020,XENON:2022ltv} and astrophysical constraints~\cite{ViauxNeutrinoAxionBounds2013,FerreiraDirectDetectionExperiments2022,AnNewStellarConstraints2013,RoachNuSTARTestsSterileneutrino2020}, with currently excluded parameter space shaded in light gray.
Top left: spin-independent WIMP--nucleon cross-section when considering the Migdal effect and with quenching fluctuations (QF) for the NR signal. Center left and bottom left: light dark matter cross-section for an heavy and light mediator, respectively. Thick lines show cross-sections giving the relic DM abundance through freeze-in~\cite{Bhattiprolu:2023akk,Boddy:2024vgt} or freeze-out~\cite{Essig:2015cda} production mechanisms. Top right: axion-electron coupling strength $g_{Ae}$. Center right: kinetic mixing parameter $\kappa$ between the photon and dark photon. Bottom right: mixing angle $|U_{e4}|^2$ between sterile neutrino and an active neutrino state. For the latter, the authors of Ref.~\cite{DarkSide:2022knj}, who contributed to this work as well, identified an error in evaluating the limit, which is corrected here.
The indirect detection limits set by the NuSTAR experiment~\cite{RoachNuSTARTestsSterileneutrino2020}, which looks for anomalous X-ray lines from radiative sterile neutrino DM decays, extends downwards to $|U_{e4}|^2=10^{-13}$ at 20 keV/c$^2$.
A local dark matter density of 0.3~GeV/c$^2$/cm$^3$ is assumed.
}
\label{fig:Migdal_LDM}
\end{figure*}

The sensitivities to other models described in Refs.~\cite{DarkSide:2022dhx} and~\cite{DarkSide:2022knj} have been evaluated. They are shown in Figure~\ref{fig:Migdal_LDM} for one year of data, using both conservative and ultimate fit approaches. In both cases and for all models, significant improvements in sensitivity over \DSfs is found over the whole mass ranges, resulting in sensitivity to uncharted phase spaces. 

\section{Conclusions}
The dual-phase liquid argon time projection chamber is presently one of the leading technologies to search for light galactic dark matter particles with masses below 10~GeV/c$^2$, as demonstrated by the \DSfs experiment with 50~kg of underground liquid argon. Based on this success, a prospect analysis was carried out for the upcoming \DStw experiment which will have a 1,000 times larger LAr volume target. 
Already with one year of data, a sensitivity improvement to DM--matter interaction cross-sections by at least one order of magnitude with respect to \DSfs is expected for a wide range of \DM models: WIMP with or without the Migdal effect in the MeV/c$^2$ to GeV/c$^2$ mass range as well as Light Dark Matter, axion-like particles, sterile neutrinos and dark photons in the keV/c$^2$ or sub-keV/c$^2$ mass range. 
A sensitivity to WIMP--nucleon interaction cross-sections below $1\times10^{-42}$~cm$^2$ is achievable for WIMP masses above 800~MeV/c$^2$. With 10 years exposure, the neutrino fog in LAr with index $n=2$ can be reached for WIMP masses around 5~GeV/c$^2$.
Further improvements in the sensitivity of \DStw\ could be achieved through dedicated measurements aimed at quantifying the fluctuations in quenching for nuclear recoils or identifying the processes underlying the generation of spurious electrons, thereby enabling their effective suppression.
\\

\vspace*{0.5cm}
{\bf Data availability} 
The expected limits reported in Figs. 3-4 and Supplementary Figs. 1-3 can be found on Zenodo~\cite{zenodo}. The other data that support the findings of this study are available from the corresponding authors upon request.

\begin{acknowledgments}
\vspace*{0.5cm}
{\bf Acknowledgements} This report is based upon work supported by the U. S. National Science Foundation (NSF) (Grants No. PHY-0919363, No. PHY-1004054, No. PHY-1004072, No. PHY-1242585, No. PHY-1314483, No. PHY-1314507, No. PHY-2310091, associated collaborative grants, No. PHY-1211308, No. PHY-1314501, No. PHY-1455351 and No. PHY-1606912, as well as Major Research Instrumentation Grant No. MRI-1429544), the Italian Istituto Nazionale di Fisica Nucleare (Grants from Italian Ministero dell’Istruzione, Università, e Ricerca Progetto Premiale 2013 and Commissione Scientific Nazionale II), the Natural Sciences and Engineering Research Council of Canada, SNOLAB, and the Arthur B. McDonald Canadian Astroparticle Physics Research Institute.\\
This work received support from the French government under the France 2030 investment plan, as part of the Excellence Initiative of Aix-Marseille University -- A*MIDEX (AMX-19-IET-008 -- IPhU).
We also acknowledge the financial support by LabEx UnivEarthS (ANR-10-LABX-0023 and ANR18-IDEX-0001), Chinese Academy of Sciences (113111KYSB20210030) and National Natural Science Foundation of China (12020101004).
This work has been supported by the S\~{a}o Paulo Research Foundation (FAPESP) grant 2021/11489-7. I. Albuquerque and  E.M. Santos are
partially supported by the National Council for Scientific and Technological Development (CNPq).
The authors were also supported by the Spanish Ministry of Science and Innovation (MICINN) through the grant PID2019-109374GB-I00, the ``Atraccion de Talento'' grant 2018-T2/TIC-10494, the Polish NCN (Grant No. UMO-2019/33/B/ST2/02884), the Polish Ministry of Science and Higher Education (MNiSW, grant number 6811/IA/SP/2018), the International Research Agenda Programme AstroCeNT (Grant No. MAB/2018/7) funded by the Foundation for Polish Science from the European Regional Development Fund, the European Union’s Horizon 2020 research and innovation program under grant agreement No 952480 (DarkWave), the Science and Technology Facilities Council, part of the United Kingdom Research and Innovation, and The Royal Society (United Kingdom), and IN2P3-COPIN consortium (Grant No. 20-152). We also wish to acknowledge the support from Pacific Northwest National Laboratory, which is operated by Battelle for the U.S. Department of Energy under Contract No. DE-AC05-76RL01830.
This research was supported by the Fermi National Accelerator Laboratory (Fermilab), a U.S. Department of Energy, Office of Science, HEP User Facility. Fermilab is managed by Fermi Research Alliance, LLC (FRA), acting under Contract No. DE-AC02-07CH11359.\\
\end{acknowledgments}

\clearpage

\bibliography{LowMass}

\clearpage

\appendix
\setcounter{figure}{0}
\renewcommand{\figurename}{Supplementary FIG.}

\section{Supplementary material}
\label{sec:NQ}

\begin{figure}[th]
\centering
\includegraphics[width=\columnwidth]{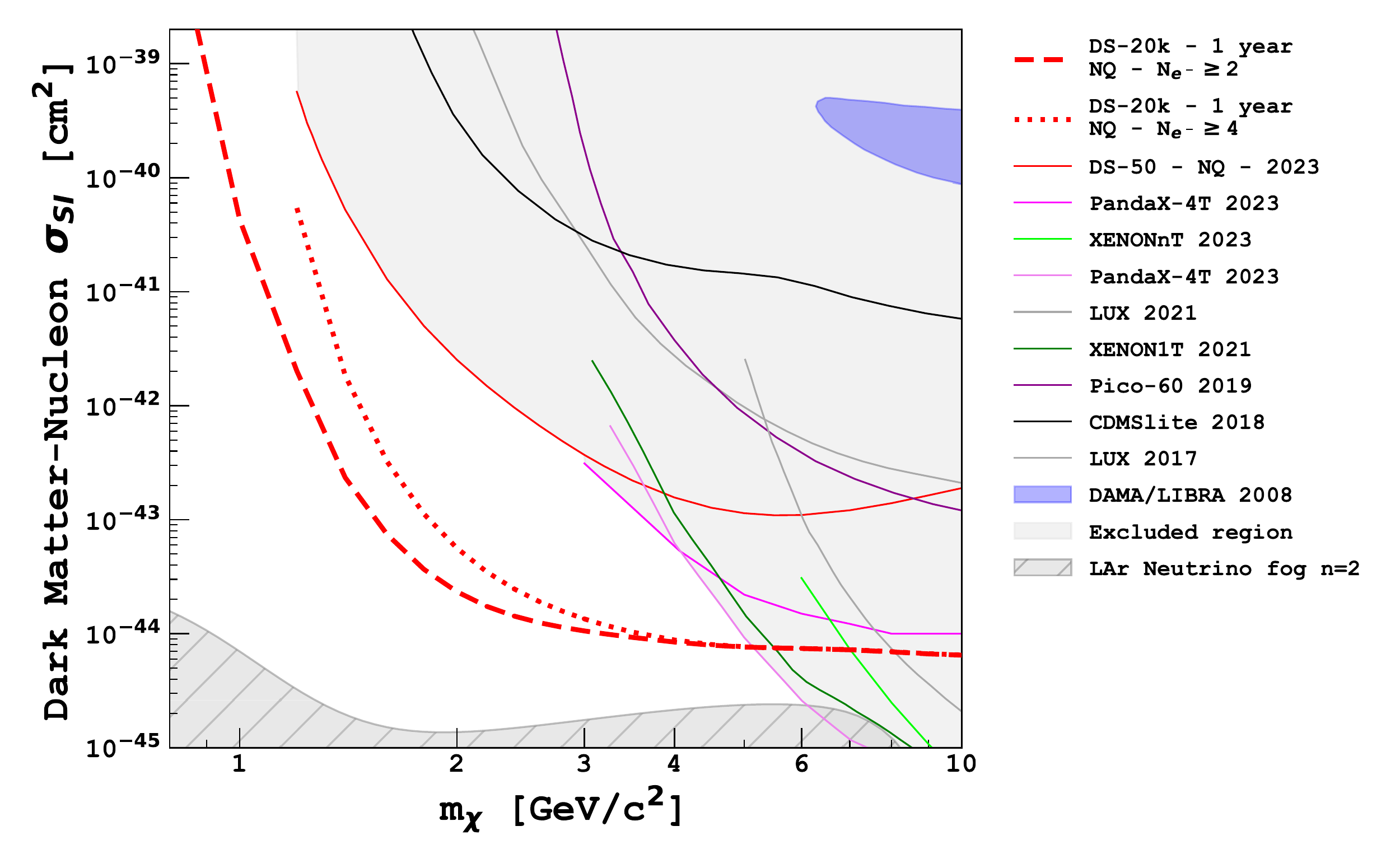}
\caption{Expected \DStw 90\% C.L. exclusion limits for spin-independent WIMP NR without quenching fluctuations (NQ) are shown as bold red lines (dotted: fit from \Ne=4, dashed: fit from \Ne=2). One year of data is assumed. They are compared to the published 90\% C.L. limits from \DSfs~\cite{DarkSide-50:2022qzh} and from other experiments~\cite{XENON:2023cxc,PandaX:2022aac,PandaX:2022xqx,XENON:2020gfr,LUX:2020yym,LUX:2016ggv,SuperCDMS:2017nns,PICO:2019vsc}, with currently excluded parameter space shaded in light gray, as well as claimed discovery from Ref.\cite{Bernabei:2013xsa}. The neutrino fog in LAr with index $n=2$~\cite{OHare:2021utq} is also shown. A local dark matter density of 0.3~GeV/c$^2$/cm$^3$ is assumed.} 
\label{fig:WIMP_annex}
\end{figure}

\begin{figure}[th]
\centering
\includegraphics[width=\columnwidth]{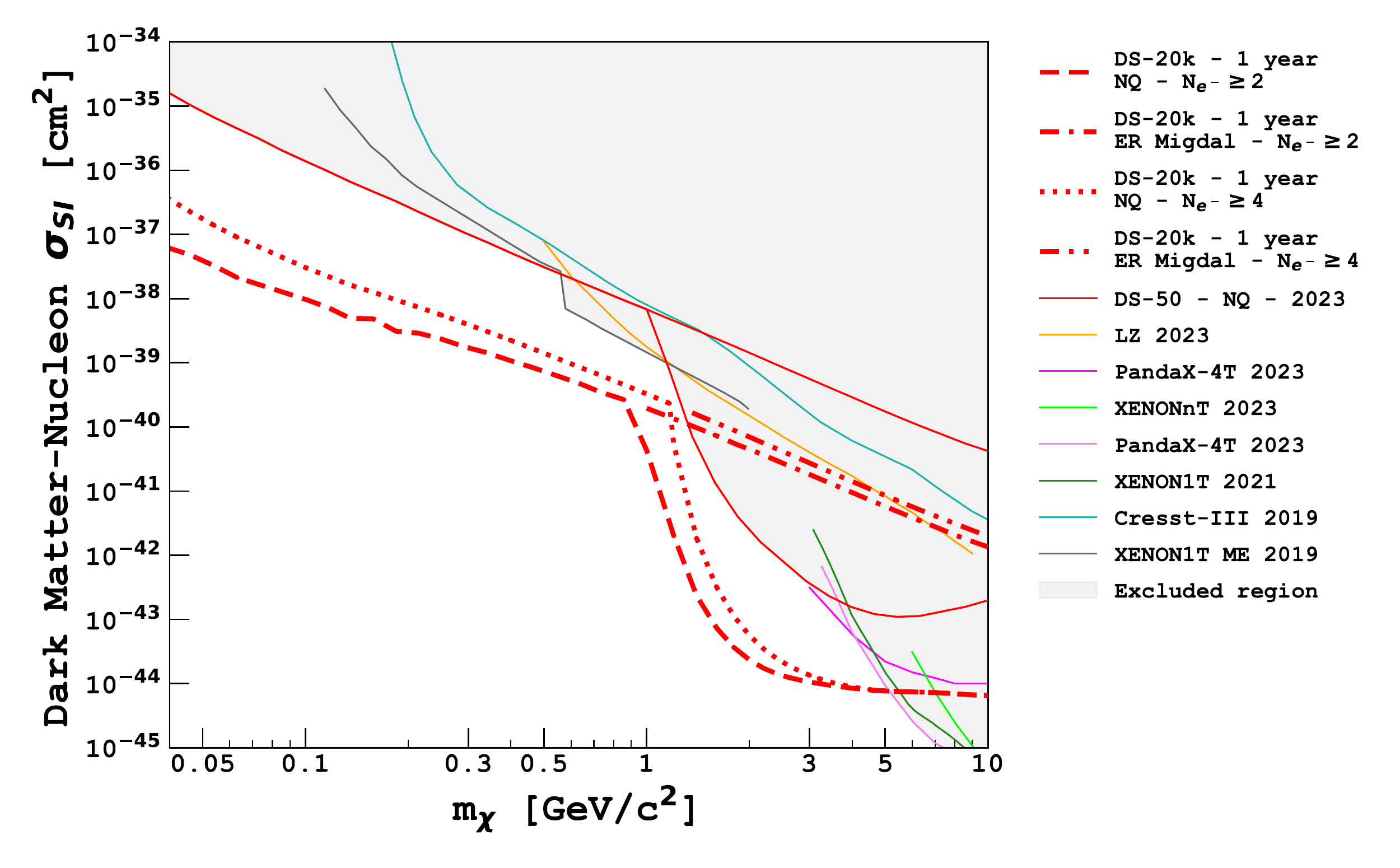}
\caption{Expected \DStw upper limits at 90\% C.L. for 
spin-independent WIMP--nucleon cross-section when considering the Migdal effect and without quenching fluctuations (NQ) for the NR signal (bold red lines, dotted: fit from \Ne=4, dashed: fit from \Ne=2). One year of data is assumed. These results are compared to the published 90\% C.L. limits from \DSfs~\cite{DarkSide:2022dhx} and other experiments~\cite{XENON:2023cxc,PandaX:2022aac,PandaX:2022xqx,XENON:2019zpr,XENON:2020gfr,PhysRevD.100.102002,LZ:2023poo}, with currently excluded parameter space shaded in light gray. A local dark matter density of 0.3~GeV/c$^2$/cm$^3$ is assumed.} 
\label{fig:Migdal_annex}
\end{figure}

\begin{figure}[h]
\centering
\includegraphics[width=\columnwidth]{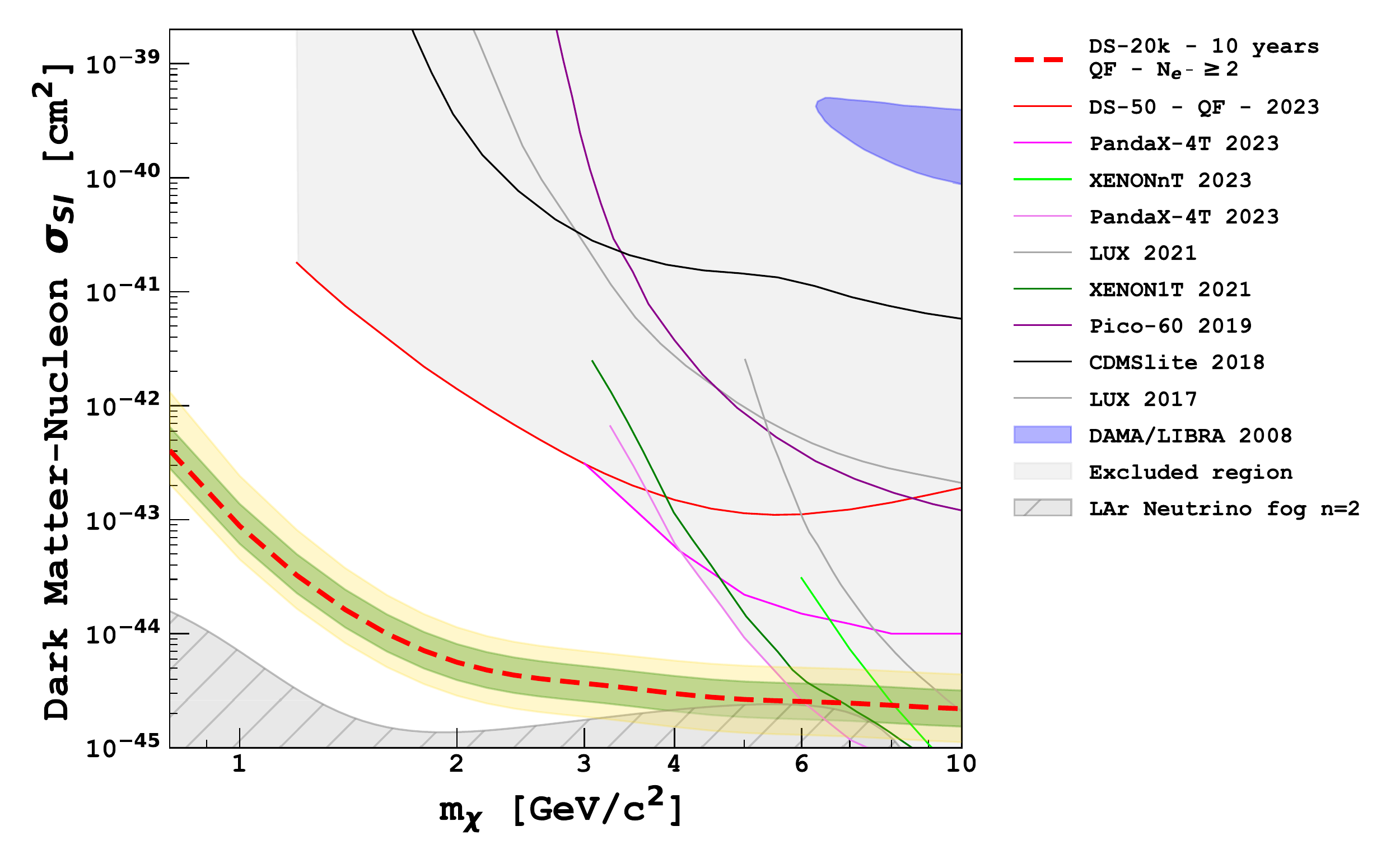}
\caption{Expected \DStw 90\% C.L. exclusion limits for spin-independent WIMP NR with quenching fluctuations (QF) are shown as bold dashed red line (fit from \Ne=2), with the $\pm$1-$\sigma$ (green shaded area) and $\pm$2-$\sigma$ (yellow
shaded area) bands. Ten years of data are assumed. They are compared to the published 90\% C.L. limits from \DSfs~\cite{DarkSide-50:2022qzh} and from other experiments~\cite{XENON:2023cxc,PandaX:2022aac,PandaX:2022xqx,XENON:2020gfr,LUX:2020yym,LUX:2016ggv,SuperCDMS:2017nns,PICO:2019vsc}, with currently excluded parameter space shaded in light gray, as well as claimed discovery from Ref.~\cite{Bernabei:2013xsa}. The neutrino fog in LAr with index $n=2$~\cite{OHare:2021utq} is also shown. A local dark matter density of 0.3~GeV/c$^2$/cm$^3$ is assumed.} 
\label{fig:WIMP_10y}

\end{figure}
\begin{figure}[h]
\centering
\includegraphics[width=\columnwidth]{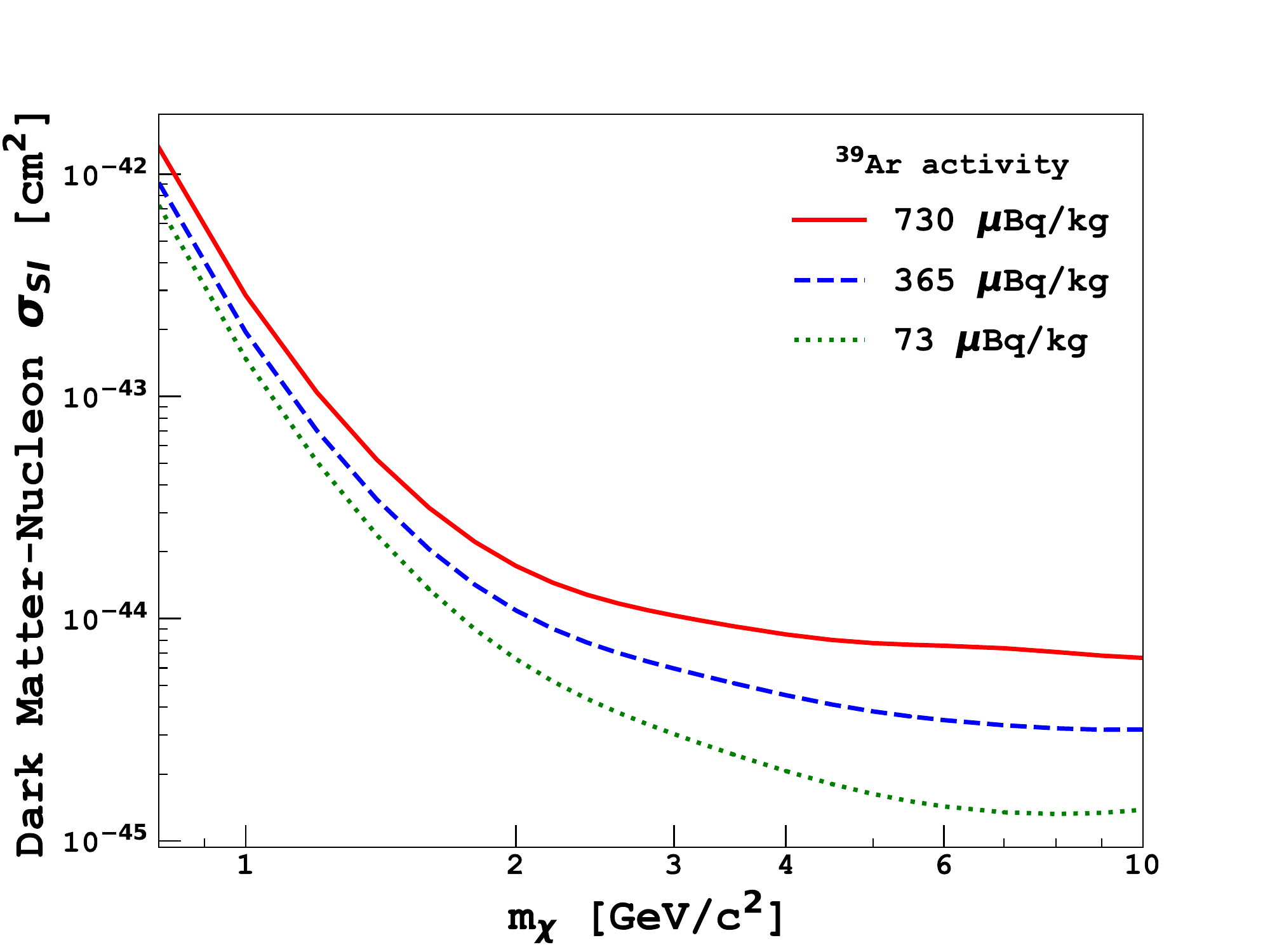}
\caption{ Expected \DStw 90\% C.L. exclusion limits for spin-independent WIMP NR with quenching fluctuations (QF) (fit from \Ne=2), for three levels of $^{39}$Ar activity: 730~$\mu$Bq$/$kg (nominal, full red line), 365~$\mu$Bq$/$kg (dashed blue line) and 73~$\mu$Bq$/$kg (dotted green line). One year of data and a local dark matter density of 0.3~GeV/c$^2$/cm$^3$ are assumed.} 
\label{fig:arlevel}
\end{figure}

\clearpage

\end{document}

%% file: authors.tex
\input{inst.tex}
\author{F.~Acerbi}\affiliation{\TNFBK}
\author{P.~Adhikari}\affiliation{\Carleton}
\author{P.~Agnes}\affiliation{\AQGSSI}\affiliation{\AQLNGS}
\author{I.~Ahmad}\affiliation{\AstroCeNT}
\author{S.~Albergo}\affiliation{\CTINFN}\affiliation{\CTUNI}
\author{I.~F.~M.~Albuquerque}\affiliation{\USP}
\author{T.~Alexander}\affiliation{\PNNL}
\author{A.~K.~Alton}\affiliation{\Augustana}
\author{P.~Amaudruz}\affiliation{\TRIUMF}
\author{M.~Angiolilli}\affiliation{\AQGSSI}\affiliation{\AQLNGS}
% ...

%% file: inst.tex
\newcommand{\Alberta}{Department of Physics, University of Alberta, Edmonton, AB T6G 2R3, Canada}
\newcommand{\APC}{APC, Universit\'e de Paris Cit\'e, CNRS, Astroparticule et Cosmologie, Paris F-75013, France}
\newcommand{\AQLNGS}{INFN Laboratori Nazionali del Gran Sasso, Assergi (AQ) 67100, Italy}
\newcommand{\AQGSSI}{Gran Sasso Science Institute, L'Aquila 67100, Italy}
\newcommand{\AQUni}{
Department of Industrial and Information Engineering and Economics, Università degli Studi dell'Aquila, L'Aquila 67100, Italy}
\newcommand{\AstroCeNT}{AstroCeNT, Nicolaus Copernicus Astronomical Center of the Polish Academy of Sciences, 00-614 Warsaw, Poland}
\newcommand{\Augustana}{Physics Department, Augustana University, Sioux Falls, SD 57197, USA}
\newcommand{\Belgorod}{Radiation Physics Laboratory, Belgorod National Research University, Belgorod 308007, Russia}
\newcommand{\BHSU}{School of Natural Sciences, Black Hills State University, Spearfish, SD 57799, USA}
\newcommand{\BINP}{Budker Institute of Nuclear Physics, Novosibirsk 630090, Russia}
\newcommand{\Birmingham}{School of Physics and Astronomy, University of Birmingham, Edgbaston, B15 2TT, Birmingham, UK}
\newcommand{\BNLaddress}{Brookhaven National Laboratory, Upton, NY 11973, USA}
\newcommand{\BOINFN}{INFN Bologna, Bologna 40126, Italy}
\newcommand{\BOUniPHY}{Department of Physics and Astronomy, Universit\`a degli Studi di Bologna, Bologna 40126, Italy}
\newcommand{\CAUniCHE}{Department of Mechanical, Chemical, and Materials Engineering, Universit\`a degli Studi, Cagliari 09042, Italy}
\newcommand{\CAUniEEE}{Department of Electrical and Electronic Engineering, Universit\`a degli Studi di Cagliari, Cagliari 09123, Italy}
\newcommand{\CAUniPHY}{Physics Department, Universit\`a degli Studi di Cagliari, Cagliari 09042, Italy}
\newcommand{\CAINFN}{INFN Cagliari, Cagliari 09042, Italy}
\newcommand{\Carleton}{Department of Physics, Carleton University, Ottawa, ON K1S 5B6, Canada}
\newcommand{\Campinas}{Physics Institute, Universidade Estadual de Campinas, Campinas 13083, Brazil}
\newcommand{\Columbia}{Physics Department, Columbia University, New York, NY 10027, USA}
\newcommand{\Chicago}{Department of Physics and Kavli Institute for Cosmological Physics, University of Chicago, Chicago, IL 60637, USA}
\newcommand{\BOCentroFermi}{Museo Storico della Fisica e Centro Studi e Ricerche Enrico Fermi, Roma 00184, Italy}
\newcommand{\CERNaddress}{CERN, European Organization for Nuclear Research 1211 Geneve 23, Switzerland, CERN}
\newcommand{\CIEMAT}{CIEMAT, Centro de Investigaciones Energ\'eticas, Medioambientales y Tecnol\'ogicas, Madrid 28040, Spain}
\newcommand{\Cluj}{National Institute for R\&D of Isotopic and Molecular Technologies, Cluj-Napoca, 400293, Romania}
\newcommand{\CPPM}{Centre de Physique des Particules de Marseille, Aix Marseille Univ, CNRS/IN2P3, CPPM, Marseille, France}
\newcommand{\CTINFN}{INFN Catania, Catania 95121, Italy}
\newcommand{\CTUNI}{Universit\`a of Catania, Catania 95124, Italy}
\newcommand{\CTLNS}{INFN Laboratori Nazionali del Sud, Catania 95123, Italy}
\newcommand{\ENSMP}{\'Ecole nationale sup\'erieure des mines de Paris, Paris 75272, France}
\newcommand{\ENUniCEE}{Engineering and Architecture Faculty, Universit\`a di Enna Kore, Enna 94100, Italy}
\newcommand{\ETHZ}{Institute for Particle Physics, ETH Z\"urich, Z\"urich 8093, Switzerland}
\newcommand{\FNAL}{Fermi National Accelerator Laboratory, Batavia, IL 60510, USA}
\newcommand{\FortLewis}{Department of Physics and Engineering, Fort Lewis College, Durango, CO 81301, USA}
\newcommand{\GEUni}{Physics Department, Universit\`a degli Studi di Genova, Genova 16146, Italy}
\newcommand{\GEINFN}{INFN Genova, Genova 16146, Italy}
\newcommand{\Hawaii}{Department of Physics and Astronomy, University of Hawai'i, Honolulu, HI 96822, USA}
\newcommand{\Houston}{Department of Physics, University of Houston, Houston, TX 77204, USA}
\newcommand{\IHEP}{Institute of High Energy Physics, Chinese Academy of Sciences, Beijing 100049, China}
\newcommand{\INFN}{Istituto Nazionale di Fisica Nucleare, Roma 00186, Italia}
\newcommand{\IPNO}{Institut de Physique Nucl\`eaire d'Orsay, 91406, Orsay, France}
\newcommand{\INSTM}{Interuniversity Consortium for Science and Technology of Materials, Firenze 50121, Italy}
\newcommand{\IPHC}{IPHC, Universit\'e de Strasbourg, CNRS/IN2P3, Strasbourg 67037, France}
\newcommand{\JINR}{Joint Institute for Nuclear Research, Dubna 141980, Russia}
\newcommand{\Krakow}{M.~Smoluchowski Institute of Physics, Jagiellonian University, 30-348 Krakow, Poland}
\newcommand{\Kurchatov}{National Research Centre Kurchatov Institute, Moscow 123182, Russia}
\newcommand{\Laurentian}{Department of Physics and Astronomy, Laurentian University, Sudbury, ON P3E 2C6, Canada}
\newcommand{\Lancaster}{Physics Department, Lancaster University, Lancaster LA1 4YB, UK}
\newcommand{\Liverpool}{Department of Physics, University of Liverpool, The Oliver Lodge Laboratory, Liverpool L69 7ZE, UK}
\newcommand{\LNFINFN}{INFN Laboratori Nazionali di Frascati, Frascati 00044, Italy}
\newcommand{\LNLINFN}{INFN Laboratori Nazionali di Legnaro, Legnaro (Padova) 35020, Italy}
\newcommand{\Lodz}{Institute of Applied Radiation Chemistry, Lodz University of Technology, 93-590 Lodz, Poland}
\newcommand{\LPNHE}{LPNHE, CNRS/IN2P3, Sorbonne Universit\'e, Universit\'e Paris Diderot, Paris 75252, France}
\newcommand{\Mainz}{Institut f\"ur Kernphysik, Johannes Gutenberg-Universit\"at Mainz, D-55128 Mainz, Germany}
\newcommand{\Manchester}{Department of Physics and Astronomy, The University of Manchester, Manchester M13 9PL, UK}
\newcommand{\MEPhI}{National Research Nuclear University MEPhI, Moscow 115409, Russia}
\newcommand{\MendeleevUniverisity}{Mendeleev University of Chemical Technology, Moscow 125047, Russia}
\newcommand{\MIBIINFN}{INFN Milano Bicocca, Milano 20126, Italy}
\newcommand{\MIINFN}{INFN Milano, Milano 20133, Italy}
\newcommand{\MIPoliICA}{Civil and Environmental Engineering Department, Politecnico di Milano, Milano 20133, Italy}
\newcommand{\MIPoliCHE}{Chemistry, Materials and Chemical Engineering Department ``G.~Natta", Politecnico di Milano, Milano 20133, Italy}
\newcommand{\MIPoliEIB}{Electronics, Information, and Bioengineering Department, Politecnico di Milano, Milano 20133, Italy}
\newcommand{\MIPoliENE}{Energy Department, Politecnico di Milano, Milano 20133, Italy}
\newcommand{\MIUni}{Physics Department, Universit\`a degli Studi di Milano, Milano 20133, Italy}
\newcommand{\MSU}{Skobeltsyn Institute of Nuclear Physics, Lomonosov Moscow State University, Moscow 119234, Russia}
\newcommand{\NAINFN}{INFN Napoli, Napoli 80126, Italy}
\newcommand{\NAUniPHY}{Physics Department, Universit\`a degli Studi ``Federico II'' di Napoli, Napoli 80126, Italy}
\newcommand{\NAUniCHE}{Chemical, Materials, and Industrial Production Engineering Department, Universit\`a degli Studi ``Federico II'' di Napoli, Napoli 80126, Italy}
\newcommand{\NAUniDIST}{Department of Structures for Engineering and Architecture, Universit\`a degli Studi ``Federico II'' di Napoli, Napoli 80126, Italy}
\newcommand{\NAUniPHARM}{Pharmacy Department, Universit\`a degli Studi ``Federico II'' di Napoli, Napoli 80131, Italy}
\newcommand{\NAUniStruct}{Department of Strutture per l'Ingegneria e l'Architettura, Universit\`a degli Studi ``Federico II'' di Napoli, Napoli 80131, Italy}
\newcommand{\NAUniEEIT}{Department of Electrical Engineering and Information Technology, Universit\`a degli Studi ``Federico II'' di Napoli, Napoli 80125, Italy}
\newcommand{\NSU}{Novosibirsk State University, Novosibirsk 630090, Russia}
\newcommand{\OACINAF}{INAF Osservatorio Astronomico di Capodimonte, 80131 Napoli, Italy}
\newcommand{\Oxford}{University of Oxford, Oxford OX1 2JD, United Kingdom}
\newcommand{\Petersburg}{Saint Petersburg Nuclear Physics Institute, Gatchina 188350, Russia}
\newcommand{\PGUniCBB}{Chemistry, Biology and Biotechnology Department, Universit\`a degli Studi di Perugia, Perugia 06123, Italy}
\newcommand{\PGINFN}{INFN Perugia, Perugia 06123, Italy}
\newcommand{\PIINFN}{INFN Pisa, Pisa 56127, Italy}
\newcommand{\PIUniPHY}{Physics Department, Universit\`a degli Studi di Pisa, Pisa 56127, Italy}
\newcommand{\PNNL}{Pacific Northwest National Laboratory, Richland, WA 99352, USA}
\newcommand{\Princeton}{Physics Department, Princeton University, Princeton, NJ 08544, USA}
\newcommand{\Queens}{Department of Physics, Engineering Physics and Astronomy, Queen's University, Kingston, ON K7L 3N6, Canada}
\newcommand{\RHUL}{Department of Physics, Royal Holloway University of London, Egham TW20 0EX, UK}
\newcommand{\RMTreINFN}{INFN Roma Tre, Roma 00146, Italy}
\newcommand{\RMTreUni}{Mathematics and Physics Department, Universit\`a degli Studi Roma Tre, Roma 00146, Italy}
\newcommand{\RMUnoINFN}{INFN Sezione di Roma, Roma 00185, Italy}
\newcommand{\RMUnoUni}{Physics Department, Sapienza Universit\`a di Roma, Roma 00185, Italy}
\newcommand{\SAINFN}{INFN Salerno, Salerno 84084, Italy}
\newcommand{\SNL}{Savannah River National Laboratory, Jackson, SC 29831, United States}
\newcommand{\SAUni}{Physics Department, Universit\`a degli Studi di Salerno, Salerno 84084, Italy}
\newcommand{\SNOLAB}{SNOLAB, Lively, ON P3Y 1N2, Canada}
\newcommand{\SSUniCHP}{Chemistry and Pharmacy Department, Universit\`a degli Studi di Sassari, Sassari 07100, Italy}
\newcommand{\STFCInterconnect}{Science \& Technology Facilities Council (STFC), Rutherford Appleton Laboratory, Technology, Harwell Oxford, Didcot OX11 0QX, UK}
\newcommand{\Sussex}{Physics and Astronomy Department, University of Sussex, Brighton BN1 9QH, UK}
\newcommand{\Temple}{Physics Department, Temple University, Philadelphia, PA 19122, USA}
\newcommand{\TNFBK}{Fondazione Bruno Kessler, Povo 38123, Italy}
\newcommand{\TNTIFPA}{Trento Institute for Fundamental Physics and Applications, Povo 38123, Italy}
\newcommand{\TNUni}{Physics Department, Universit\`a degli Studi di Trento, Povo 38123, Italy}
\newcommand{\TOINFN}{INFN Torino, Torino 10125, Italy}
\newcommand{\TOPoli}{Department of Electronics and Communications, Politecnico di Torino, Torino 10129, Italy}
\newcommand{\TOUni}{Physics Department, Universit\`a degli Studi di Torino, Torino 10125, Italy}
\newcommand{\TRIUMF}{TRIUMF, 4004 Wesbrook Mall, Vancouver, BC V6T 2A3, Canada}
\newcommand{\TUM}{Physik Department, Technische Universit\"at M\"unchen, Munich 80333, Germany}
\newcommand{\UB}{Universiatat de Barcelona, Barcelona E-08028, Catalonia, Spain} 
\newcommand{\UCDavis}{Department of Physics, University of California, Davis, CA 95616, USA}
\newcommand{\UCRiverside}{Department of Physics and Astronomy, University of California, Riverside, CA 92507, USA}
\newcommand{\UCSanDiego}{Department of Physics, University of California, San Diego, CA 92093, USA}
\newcommand{\UCLA}{Physics and Astronomy Department, University of California, Los Angeles, CA 90095, USA}
\newcommand{\UCAS}{University of Chinese Academy of Sciences, Beijing 100049, China}
\newcommand{\UMass}{Amherst Center for Fundamental Interactions and Physics Department, University of Massachusetts, Amherst, MA 01003, USA}
\newcommand{\UNAM}{Instituto de F\'isica, Universidad Nacional Aut\'onoma de M\'exico, M\'exico 01000, Mexico}
\newcommand{\UnivAQ}{Universit\`a degli Studi dell’Aquila, L’Aquila 67100, Italy}
\newcommand{\UniversityofEdinburgh}{School of Physics and Astronomy, University of Edinburgh, Edinburgh EH9 3FD, UK}
\newcommand{\UOC}{Department of Chemistry, University of Crete, P.O. Box 2208, 71003 Heraklion, Crete, Greece}
\newcommand{\USP}{Instituto de F\'isica, Universidade de S\~ao Paulo, S\~ao Paulo 05508-090, Brazil}
\newcommand{\VTech}{Virginia Tech, Blacksburg, VA 24061, USA}
\newcommand{\Warwick}{University of Warwick, Department of Physics, Coventry CV47AL, UK}
\newcommand{\WUT}{Institute of Radioelectronics and Multimedia Technology, Warsaw University of Technology, 00-661 Warsaw, Poland}
\newcommand{\WilliamsCollege}{Williams College, Physics Department, Williamstown, MA 01267 USA}
\newcommand{\Zaragoza}{Centro de Astropart\'iculas y F\'isica de Altas Energ\'ias, Universidad de Zaragoza, Zaragoza 50009, Spain}
\newcommand{\ZaragozaARAID}{Fundaci\'on ARAID, Universidad de Zaragoza, Zaragoza 50009, Spain}
\newcommand{\UniHAM}{Institute of Experimental Physics, University of Hamburg, Luruper Chaussee 149, 22761, Hamburg, Germany}